\documentclass[12pt,preprint]{aastex}
\def\lea{\mathrel{<\kern-1.0em\lower0.9ex\hbox{$\sim$}}}
\def\gea{\mathrel{>\kern-1.0em\lower0.9ex\hbox{$\sim$}}}

\slugcomment{Submitted to The Astrophysical Journal}

\shorttitle{Faint UV Standards}
\shortauthors{Siegel et al.}

\begin{document}

\title{Faint NUV/FUV Standards from Swift/UVOT, GALEX and SDSS Photometry}

\author{Michael H. Siegel\altaffilmark{1}, Erik A. Hoversten\altaffilmark{1}, Peter W. A. Roming\altaffilmark{1},
Wayne B. Landsman\altaffilmark{2}, Carlos Allende Prieto\altaffilmark{3,4}, Alice A. Breeveld\altaffilmark{3},
Peter Brown\altaffilmark{1,5}, Stephen T. Holland\altaffilmark{2}, N. P. M. Kuin\altaffilmark{3}, Mathew J. Page\altaffilmark{3}.
Daniel E. Vanden Berk\altaffilmark{6}}

\altaffiltext{1}{Pennsylvania State University, Department of Astronomy, 525 Davey Laboratory, University Park, PA, 16802 (siegel@astro.psu.edu,
hoversten@astro.psu.edu, roming@astro.psu.edu, brown@astro.psu.edu)}
\altaffiltext{2}{NASA/Goddard Space Flight Center, Astrophysics Science Division, Code 661, Greenbelt, MD 20771
(wayne.b.landsman@nasa.gov, Stephen.T.Holland@nasa.gov) }
\altaffiltext{3}{Mullard Space Science Laboratory, University College London, Holmbury St. Mary, Dorking, Surrey RH5 6NT
(cap@mssl.ucl.ac.uk, aab@mssl.ucl.ac.uk, npmk@mssl.ucl.ac.uk, mjp@mssl.ucl.ac.uk)}
\altaffiltext{4}{Current address: Instituto de Astrof\'{\i}sica de Canarias,
V\'{\i}a L\'actea s/n, E-38205 La Laguna, Tenerife, Spain}
\altaffiltext{5}{Current address: Department of Physics and Astronomy, University of Utah, Salt Lake City, UT 84112}
\altaffiltext{6}{Physics Department, St. Vincent College, Latrobe, PA 15650 (daniel.vandenberk@email.stvincent.edu)}
	
\begin{abstract}
At present, the precision of deep ultraviolet photometry is somewhat limited by the dearth of faint ultraviolet standard stars.
In an effort to improve this situation, we present a uniform catalog of eleven new faint $(u\sim17$) ultraviolet
standard stars.  High-precision photometry of these stars has been taken from the Sloan Digital Sky Survey and
{\it Galaxy Evolution Explorer}
and combined with new data from the {\it Swift} Ultraviolet Optical Telescope to provide precise photometric measures
extending from the Near Infrared
to the Far Ultraviolet.  These stars were chosen because they are known to be hot ($20,000 < T_{eff} < 50,000 K$) DA white dwarfs with published
Sloan spectra that should be photometrically stable.  This careful selection allows us to compare
the combined photometry and Sloan spectroscopy to models of pure hydrogen atmospheres 
to both constrain the underlying properties of the white dwarfs and test the ability of white dwarf models
to predict the photometric measures.  We find that the photometry provides good constraint on white dwarf temperatures, which
demonstrates the ability of Swift/UVOT to investigate the properties of hot luminous stars.  We further find that the 
models reproduce the photometric
measures in all eleven passbands to within their systematic uncertainties.  Within the limits of our photometry, we find
the standard stars to be photometrically stable.
This success indicates that the 
models can be used to calibrate additional filters to our standard system, permitting easier
comparison of photometry from heterogeneous sources.  The largest source of uncertainty in the model fitting is
the uncertainty in the foreground reddening curve, a problem that is especially acute in the UV.
\end{abstract}


\section{Introduction}
\label{s:intro}

The last three decades have witnessed the advent of numerous space-based ultraviolet-sensitive instruments.
Programs such as the {\it Hubble Space Telescope} Faint Object
Spectrograph (FOS), Space Telescope Imaging Spectrograph (STIS), and Advanced Camera for Surveys
(ACS), {\it International Ultraviolet Explorer}, {\it Far Ultraviolet Spectroscopic Explorer}, {\it Galex
Evolution Explorer} and {\it Hopkins Ultraviolet Telescope} have created an infusion of
scientific discovery, particularly in hot or high-energy environments.

The expansion of ultraviolet astronomy, however, has run into a problem of calibration.
The primary set of UV calibration standards for the above missions consists of four hot white dwarf stars --
G 191-B2B, GD 153, GD 71, and HZ 43 (Bohlin 1996, 2000, 2007; Bohlin et al. 2001; Bohlin \& Gilliland 2004; 
Nichols \& Linsky 1996; Kruk et al. 1999).
All four, however, are brighter than $m_V=13.4$ (Holberg and Bergeron 2006).  While 
such bright standards were excellent
for previous generations of instruments, they are too bright for the latest generation of telescopes.
The Bohlin standards would quickly saturate CCD cameras on large telescopes
(or, in the case of the Cosmic Origins Spectrograph, potentially damage the detector) and short exposure times
bring shutter resolution into play.
Observations with photon-counting
instruments -- such as the {\it Swift} Ultraviolet
Optical Telescope (UVOT), ASTROSAT's Ultraviolet
Imaging Telescope or the Tel Aviv University UV Explorer -- are compromised by
coincidence loss.  Coincidence loss occurs
when two or more photons from an astronomical source arrive within a single detector read time and are
therefore read as a single photon \citep{Fordham00}.
The brighter the source, the greater the coincidence
loss.  Coincidence loss can not be ameliorated by shorter exposure times.
Beyond a certain range (with UVOT, about 100 ph sec$^{-1}$) coincidence loss 
becomes so great as to make measured brightnesses unreliable (see Poole et al. 2008, hereafter P08, their figure 6).

Recent calibration efforts using faint UV standards have been made by the {\it Swift}/UVOT team 
(P08).  However, even UVOT is only
calibrated to three objects -- WD 1657+343, WD 1121+145 and WD 1026+453 -- in the UV passbands.  These hot white dwarfs have $U$ magnitudes
of 14.8-15.4 and 
were observed as part of an HST faint extension calibration program (10094).  However, the HST program
was unable to proceed after 2003 owing to the failure of STIS and a fourth faint standard (WD 0947+857) was suspected
to have a composite UV-optical spectrum (Lajoie \& Bergeron, 2007).
The need for a larger number of faint UV standards
remains critical.

\begin{center}
\begin{figure}[ht!]
\includegraphics[angle=90, scale=.65]{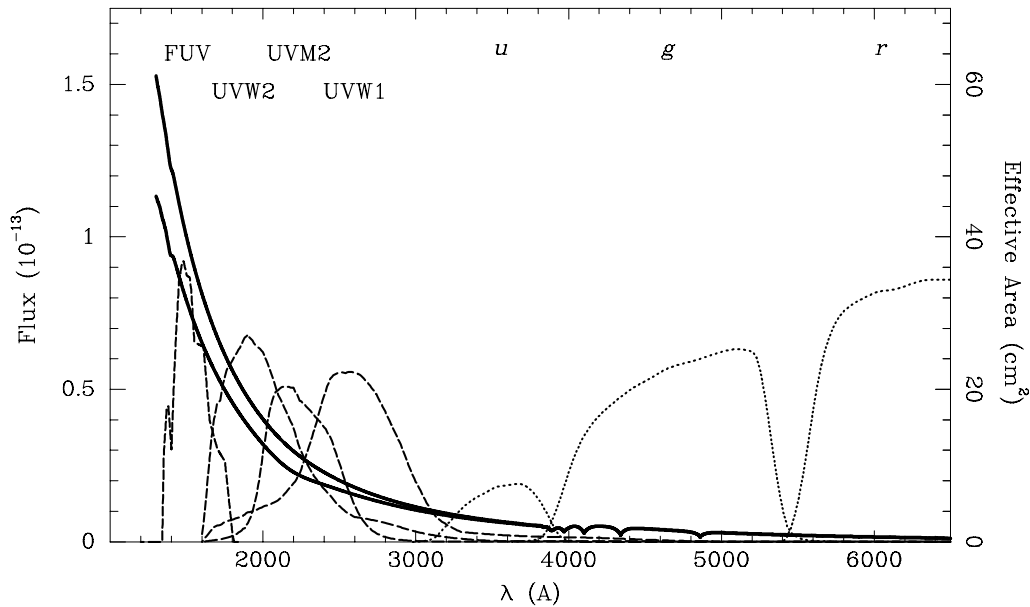}
\figcaption[f1.eps]{The sensitivity of the NUV and FUV passbands to the properties of white dwarfs.  The thick lines are two of our fitted
white dwarf models.  The upper line is SDSS J150050.71+040430.0; the lower line is SDSS J173020.12+613937.5.  Note that the flux is dominated
by the UV emission, where the {\it GALEX} and {\it Swift}/UVOT filters (dashed lines) are centered.  Note also the lack of distinct spectral lines
in the UV passbands.  The SDSS filters are shown in dotted lines
and arbitrarily scaled down to provide a comparison to the NUV and FUV filters.\label{f:speclook}}
\end{figure}
\end{center}

A recent study by Allende Prieto et al. (2009, hereafter AP09) has taken the first step in this direction.  Using data from the
Sloan Digital Sky Survey (SDSS), they identify nine hot faint DA white dwarfs as potential
spectrophotometric standards.
Hot DA white dwarfs are suitable as UV standard candles because of their high UV
luminosity, blue colors and the paucity of spectral lines that makes them easy to model.  In turn, because the
UV passbands are sensitive to the properties of white dwarfs (see Figure \ref{f:speclook}), studying
white dwarf standards can provide a reciprocal test on the white dwarf models themselves and provide additional constraint on the properties of white dwarfs.

Our study complements the AP09 study by using {\it Swift}/UVOT to observe eleven faint hot DA white
dwarfs selected not only from the SDSS but also from the {\it Galaxy Evolution Explorer} (GALEX) catalog.
We use the combined data, covering
the spectral range from the NIR to the FUV, to provide tight constraints on the temperatures of the white dwarf stars.
Moreover, we test the ability of pure hydrogen models of white dwarfs to reproduce
the measured photometry.  The result is a group of 
stars with calibrated observations
in eleven passbands, published SDSS spectra and well-constrained model spectra that can be used to calibrate existing or
future instruments that may use different filters.

\section{Observations and Data}
\label{s:obsred}

\subsection{Sample Selection}
\label{ss:selection}

The first step in constructing the catalog of faint UV standards was
to select good candidate stars.  DA white dwarf stars are ideal
targets for a number of reasons.  First, while some may suffer from metal pollution, 
they are expected to manifest mostly
hydrogen absorption lines, none of which would be 
in the 1700--3000 \AA\ wavelength range of UVOT's UV filters.
This simplifies the modeling.
Second, white dwarfs have been successfully
modeled at the 1\% level over the temperature range of 20,000 K $< T_{eff}<$ 90,000 K, allowing confident comparison between theoretical
models and empirical magnitudes \citep{Holberg06}.  Third, trigonometric parallaxes have
confirmed the utility of photometric parallaxes (Holberg et al. 2008).
Fourth, outside of the
known instability strip, 
DA white dwarf luminosity variations are driven by radiative
cooling over long timescales and they are 
therefore expected to show little photometric variability.
Finally, large catalogs of spectroscopically confirmed white dwarf stars have
already been compiled (e.g., McCook \& Sion 1999; Eisenstein et al. 2006).

We began our selection with the
catalog of 9316 spectroscopically confirmed white dwarfs from the SDSS
fourth data release \citep{Eisenstein06}.  The advantage of the \citet{Eisenstein06}
catalog is that it contains uniform, high quality photometry in five
filters ranging from roughly 3500 to 10000 \AA\ as well as uniform
spectroscopy at $R\simeq 1800$ covering a wavelength range of 3800 to 9200 \AA\ \citep{DR1}.
\cite{Eisenstein06} provide spectral classifications for
each object as well as a homogenous set of effective temperatures and surface gravities.
The primary drawback of the SDSS
catalog is that the survey footprint primarily covers declinations above zero that are accessible from Apache
Point and avoids the Galactic
plane.  These limitations preclude a uniform sky distribution of standard stars.

We selected white dwarfs from the SDSS catalog that were spectroscopically
classified as DA white dwarfs.  To ensure their suitability for simple modeling,
we further restricted the sample to
stars that lack K or M star companions, strong magnetic contributions, or
evidence of helium, carbon, or other metal lines.  Additionally, 
to minimize the effects
of coincidence loss, we required that the flux in each filter
be less than 5 counts s$^{-1}$, which corresponds to a magnitude range
of $uvw2<15$, with UVOT magnitudes estimated from 
the SDSS $u$-band magnitude and the effective temperatures of
\cite{Eisenstein06}.  At count rates of 5 counts s$^{-1}$ or below, the coincidence loss is less than 3\% and 
is corrected by the formulation given in P08 to better than 1\%.

We selected
stars with stellar temperatures of 20,000 K $<T_{eff}<$ 50,000 K.
Previous investigations (e.g., Holberg \& Bergeron, 2006) have shown
DA white dwarfs in this temperature range to be modellable to better
than 1\% precision.  The lower
temperature bound also avoids the instability strip that white dwarfs cross
as they undergo radiative cooling and become DAV variable stars 
\citep[e.g.][]{
Mukadam04,Mullally05}.
We also removed stars that either had large uncertainties in
effective
temperature or surface gravity ($>$1000 K and 0.3 dex, respectively)
or where the dust maps of \cite{SFD} indicate a reddening of
$E(B-V) > 0.05$.  The latter cut is especially critical
in the ultraviolet, where extinction is much higher and more uncertain than in the optical
\citep[e.g. see][for dust models from the Milky Way and Magellanic
Clouds]{Pei92}, significantly multiplying the impact of any uncertainty in the
foreground extinction.  Moreover, the extinction law itself may vary
from $R_V\sim3.1$ over a range of 2.2-5.8, depending on the line of sight
through the Galaxy \citep{Fitzpatrick99}.

All of these cuts reduced our sample to 136 candidate stars (1.5\% of the
sample).  We then imposed the additional requirement that data for each star
be available from the GALEX mission (Martin et al. 2005; Morrissey et al. 2007).
GALEX has a near-UV filter 
with an effective wavelength of 2267 \AA, which overlaps the UVOT filters.  More importantly, it
has a far-UV filter
with an effective wavelength of 1516 \AA, bluer than the Swift filters, which
allows for even more rigorous constraint of $T_{eff}$.
While GALEX's All-Sky Imaging Survey covers most of the sky
with an exposure time of at least 100 seconds, the Medium
Imaging Survey (MIS) covers 1000 square degrees in the SDSS footprint with
exposure times greater than 1500 seconds.
We selected only stars that were covered by MIS where the
GALEX photometry error is dominated by systematics.  
The MIS
requirement eliminated most remaining candidate stars because the MIS
only covers one sixth of DR4's 6670 square degree footprint \citep{DR4}.
\footnote{The list of targets was
created in early 2008 before the release of GALEX DR5, so
this requirement would be less stringent today.}

From the remaining candidates, we selected 12 stars that were as equally spaced across
the sky as possible, given the constraint of the SDSS/MIS footprint, had no other sources within 15'' and no bright
stars within the 17' UVOT field of view (FOV).  Eleven of these were subsequently observed by Swift/UVOT and these eleven
comprise our new catalog of standards.  The final list of target
stars and observations is shown in Table \ref{t:targets}.  We list the
SDSS identification, coordinates and SDSS $u$ magnitude as well as the
number of {\it Swift} observations and total UVOT exposure time as
detailed in the next subsection. We also list the reddening values derived from the
maps of \cite{SFD}.
SDSS and GALEX photometry are listed in 
Tables \ref{t:photsdss} and \ref{t:photspace}, respectively.
In all cases, we have listed, in the top row, the uncertainty in the
photometric zero points specified
by Ivezi{\'c} et al. 2004 (SDSS), Morrissey et al. 2007 (GALEX) and P08 (Swift/UVOT).  These uncertainties
are added to the random photometric uncertainties for our analysis.

Our selection was made prior to the publication of AP09 and has only one target in common (J134430.11+032423.2).
However, AP09 had a
brighter magnitude limit, which would have excluded all but four of our target stars.  They further
refined the sample based on the quality of agreement between observations and models.  Their final
selection of nine stars, against which we have no overlap, is based on expected uncertainties in
photometry.  Our selection, by contrast, was aimed at finding stars that would produce high-quality UVOT data
and had extant
GALEX photometry.  However, Swift/UVOT observation of the AP09 standards is highly recommended.

\begin{center}
\begin{deluxetable}{l|ccc|c|cc}
\tabletypesize{\footnotesize}
\tablewidth{0 pt}
\tablecaption{Swift/GALEX/SDSS UV Standard Stars\label{t:targets}}
\tablehead{
\colhead{Target} &
\colhead{RA} &
\colhead{DEC} &
\colhead{E (B-V)$_{SFD}$} &
\colhead{$u$} &
\colhead{$N\tablenotemark{a}_{obs,Swift}$} &
\colhead{Exp Time (ks)}}
\startdata
\hline
SDSS J002806.49+010112.2 & 7.0271   & 1.0200   & 0.021 & 17.5 & 3 (70)  & 12.7\\
SDSS J083421.23+533615.6 & 128.5883 & 53.6042  & 0.039 & 16.5 & 3 (120) & 19.8\\
SDSS J092404.84+593128.8 & 141.0200 & 59.5247  & 0.025 & 17.5 & 3 (80)  & 18.5\\
SDSS J103906.00+654555.5 & 159.7750 & 65.7653  & 0.018 & 17.7 & 2 (71)  & 13.5\\
SDSS J134430.11+032423.2 & 206.1254 & 3.4064   & 0.026 & 16.5 & 1 (36)  & 4.5\\
SDSS J140641.95+031940.5 & 211.6750 & 3.3278   & 0.035 & 17.9 & 2 (64)  & 11.2\\
SDSS J144108.43+011020.0 & 220.2850 & 1.1722   & 0.040 & 16.7 & 2 (48)  & 4.4\\
SDSS J150050.71+040430.0 & 225.2113 & 4.0750   & 0.044 & 17.7 & 2 (45)  & 8.9\\
SDSS J173020.12+613937.5 & 262.5838 & 61.6603  & 0.040 & 17.8 & 4 (134) & 17.3\\
SDSS J231731.36-001604.9 & 349.3808 & -0.2681  & 0.040 & 16.4 & 3 (104) & 15.3\\
SDSS J235825.80-103413.4 & 359.6075 & -10.5703 & 0.032 & 17.3 & 3 (56)  & 11.1\\
\hline
\enddata
\tablenotetext{a}{Number of Swift epochs (Total Number of UVOT images)}
\end{deluxetable}
\end{center}

\subsection{Photometry}
\label{ss:photometry}

We supplemented the existing GALEX and SDSS 
photometry for the DA white dwarfs with a new epoch of photometry from the UVOT instrument aboard the {\it Swift}
Gamma Ray Burst Mission (Gehrels et al. 2004).  UVOT is a modified Richey-Chretien 30 cm telescope that has a 
wide (17' $\times$ 17') field of view and a microchannel plate intensified CCD operating in photon
counting mode (see details in Roming et al. 2005).
It is designed to catch the early optical/ultraviolet afterglows
of gamma ray bursts.  However, as a wide field instrument sensitive
over the wavelength range of 1700-8000 \AA\
that observes simultaneously with {\it Swift}'s X-Ray Telescope (XRT; Burrows et al. 2005), 
it fills a unique niche beyond gamma ray bursts, allowing multi-wavelength investigations into a wide range of
astrophysical phenomena.

The instrument is equipped with a filter wheel that includes a clear white filter, $u$, $b$ and $v$ optical filters, 
$uvw1$, $uvm2$ and $uvw2$ ultraviolet filters, a magnifier, two grisms and a blocked filter. The UV filters are
narrower than those of GALEX.  This narrowness limits the overall spectral range but significantly improves the spectral resolution.
For the purpose of our faint UV standard catalog, the UVOT data allow a potent
extension into the UV, providing keener sensitivity to the properties of our white dwarf standard stars, particularly their temperatures.

Eleven of our twelve target stars were observed as fill-in targets during the 2008-9 {\it Swift} AO4 observing cycle.  Data were
taken in the $u$, $uvw1$, $uvm2$ and $uvw2$ filters between June 2008 and February 2009.  A handful of
stars were re-observed in June 2009 for additional calibration.  Exposure times and sequencing varied depending
on observing windows, XRT temperature concerns and interrupting gamma ray bursts or targets of opportunity.
Multiple epochs were obtained to both
improve photometric precision and allow a check on the variability of our standard stars.

\begin{center}
\begin{deluxetable}{lccccc}
\tabletypesize{\footnotesize}
\tablewidth{0 pt}
\tablecaption{SDSS Photometry of Faint UV Standards\label{t:photsdss}}
\tablehead{
\colhead{Name} &
\colhead{$u$} &
\colhead{$g$} &
\colhead{$r$} &
\colhead{$i$} &
\colhead{$z$}}
\startdata
Systematic Uncertainty & $\pm.03$ & $\pm.01$ & $\pm.01$ & $\pm.01$ & $\pm.02$\\
\hline
SDSS J002806.49+010112.2 & $17.457\pm0.021$ & $17.559\pm0.019$ & $17.938\pm0.017$ & $18.227\pm0.020$ & $18.528\pm0.041$ \\
SDSS J083421.23+533615.6 & $16.496\pm0.029$ & $16.685\pm0.015$ & $17.088\pm0.020$ & $17.422\pm0.022$ & $17.789\pm0.031$ \\
SDSS J092404.84+593128.8 & $17.524\pm0.015$ & $17.542\pm0.032$ & $17.942\pm0.014$ & $18.276\pm0.023$ & $18.578\pm0.033$ \\
SDSS J103906.00+654555.5 & $17.729\pm0.015$ & $17.883\pm0.021$ & $18.281\pm0.018$ & $18.588\pm0.018$ & $18.870\pm0.055$ \\
SDSS J134430.11+032423.2 & $16.482\pm0.015$ & $16.603\pm0.018$ & $17.005\pm0.016$ & $17.323\pm0.016$ & $17.613\pm0.025$ \\
SDSS J140641.95+031940.5 & $17.896\pm0.024$ & $17.900\pm0.014$ & $18.300\pm0.019$ & $18.576\pm0.020$ & $18.916\pm0.051$ \\
SDSS J144108.43+011020.0 & $16.669\pm0.015$ & $16.870\pm0.020$ & $17.299\pm0.016$ & $17.604\pm0.023$ & $17.879\pm0.026$ \\
SDSS J150050.71+040430.0 & $17.744\pm0.015$ & $17.880\pm0.018$ & $18.259\pm0.014$ & $18.561\pm0.014$ & $18.793\pm0.044$ \\
SDSS J173020.12+613937.5 & $17.830\pm0.021$ & $17.837\pm0.017$ & $18.147\pm0.018$ & $18.451\pm0.019$ & $18.756\pm0.043$ \\
SDSS J231731.36-001604.9 & $16.401\pm0.020$ & $16.485\pm0.025$ & $16.834\pm0.021$ & $17.142\pm0.017$ & $17.441\pm0.023$ \\
SDSS J235825.80-103413.4 & $17.247\pm0.029$ & $17.220\pm0.029$ & $17.640\pm0.020$ & $17.880\pm0.015$ & $18.263\pm0.034$ \\
\hline
\enddata
\end{deluxetable}
\end{center}

\begin{center}
\begin{deluxetable}{lccccccc}
\tabletypesize{\tiny}
\tablewidth{0 pt}
\tablecaption{{\it Swift} UVOT and GALEX Photometry of Faint UV Standards (AB mags)\label{t:photspace}}
\tablehead{
\colhead{Name} &
\colhead{FUV} &
\colhead{NUV} &
\colhead{$uvw2$} &
\colhead{$uvm2$} &
\colhead{$uvw1$} &
\colhead{$u$} &
\colhead{Var}}
\startdata
\hline
Systematic Uncertainty & $\pm.05$   &  $\pm.03$  & $\pm.03$   & $\pm.03$      & $\pm.03$    & $\pm.02$\\
\hline
SDSS J002806.49+010112.2 & $16.446 \pm 0.004$ & $16.801 \pm 0.004$ & $16.694 \pm 0.007$ & $16.834 \pm 0.010$ & $16.977 \pm 0.010$ & $17.290 \pm 0.009$ & 0.55\\
SDSS J083421.23+533615.6 & $15.429 \pm 0.006$ & $15.893 \pm 0.004$ & $15.761 \pm 0.004$ & $15.908 \pm 0.005$ & $16.070 \pm 0.005$ & $16.431 \pm 0.005$ & 1.57\\
SDSS J092404.84+593128.8 & $16.716 \pm 0.007$ & $17.041 \pm 0.006$ & $16.979 \pm 0.007$ & $17.099 \pm 0.010$ & $17.216 \pm 0.009$ & $17.455 \pm 0.008$ & 2.45\\
SDSS J103906.00+654555.5 & $16.859 \pm 0.012$ & $17.206 \pm 0.008$ & $17.072 \pm 0.007$ & $17.197 \pm 0.010$ & $17.363 \pm 0.010$ & $17.672 \pm 0.010$ & 0.55\\
SDSS J134430.11+032423.2 & $15.434 \pm 0.007$ & $15.880 \pm 0.003$ & $15.784 \pm 0.007$ & $15.948 \pm 0.010$ & $16.118 \pm 0.010$ & $16.390 \pm 0.009$ & 0.38\\
SDSS J140641.95+031940.5 & $16.994 \pm 0.012$ & $17.433 \pm 0.009$ & $17.356 \pm 0.009$ & $17.474 \pm 0.013$ & $17.602 \pm 0.012$ & $17.819 \pm 0.013$ & 2.87\\
SDSS J144108.43+011020.0 & $15.487 \pm 0.007$ & $16.015 \pm 0.005$ & $15.882 \pm 0.008$ & $16.023 \pm 0.010$ & $16.221 \pm 0.012$ & $16.562 \pm 0.010$ & 2.02\\
SDSS J150050.71+040430.0 & $16.779 \pm 0.011$ & $17.162 \pm 0.007$ & $17.056 \pm 0.008$ & $17.223 \pm 0.015$ & $17.331 \pm 0.016$ & $17.629 \pm 0.011$ & 1.90\\
SDSS J173020.12+613937.5 & $17.069 \pm 0.010$ & $17.408 \pm 0.005$ & $17.312 \pm 0.007$ & $17.381 \pm 0.010$ & $17.486 \pm 0.010$ & $17.724 \pm 0.010$ & 1.60\\
SDSS J231731.36-001604.9 & $15.499 \pm 0.003$ & $15.893 \pm 0.002$ & $15.796 \pm 0.005$ & $15.930 \pm 0.006$ & $16.065 \pm 0.006$ & $16.314 \pm 0.005$ & 2.35\\
SDSS J235825.80-103413.4 & $16.416 \pm 0.005$ & $16.722 \pm 0.004$ & $16.662 \pm 0.008$ & $16.752 \pm 0.011$ & $16.885 \pm 0.010$ & $17.101 \pm 0.009$ & 0.79\\
\hline
\enddata
\end{deluxetable}
\end{center}

Photometry was generated and calibrated through the standard pipeline described in P08 and Marshall et al.
(in preparation).  The P08 photometric system has been
shown to be consistent at the 1-3\% level, a performance we check in \S \ref{ss:variability}.  The latest pipeline
also accounts for the 1\% per year decline in UVOT's sensitivity (Breeveld et al. 2010a).

In addition to the standard photometric transformation, we performed additional corrections which will soon
be incorporated in the UVOT calibration.  The first was a slight correction
to the zero points of P08 and revision of the $uvw1$ response curve based on observations of numerous reference stars.
This reflects a new UVOT calibration that supersedes P08 and will soon be described in Breeveld et al. (2010b, in prep).
The second was a transformation to the ABmag system.  For the P08 calibration, we transformed the Vega magnitudes to the AB
system by using the Vega spectrum of
Bohlin \& Gilliland (2004) to calculate the magnitude of Vega in the Swift filters, essentially reversing the Vegamag transformation of P08.
\footnote{The P08 zero point uncertainties quoted in Table \ref{t:photspace} include a $\sim1\%$ uncertainty arising
from the ABmag to Vegamag conversion. By effectively removing this conversion, our actual zero point
uncertainties are marginally but not significantly smaller than those given P08.}  For the revised calibration, we used
the February 2010 CALSPEC spectrum\footnote{Available at http://www.stsci.edu/hst/observatory/cdbs/calspec.html}.
The AB magnitude corrections, for both the P08 and revised
calibrations, are given in Table \ref{t:vega2ab}.

\begin{center}
\begin{deluxetable}{ccc}
\tablewidth{0 pt}
\tablecaption{Correction from Swift Vegamag to ABmag System\label{t:vega2ab}}
\tablehead{
\colhead{Filter} &
\colhead{P08 Calibration} &
\colhead{New Calibration}}
\startdata
\hline
$u$    &  1.02 & 1.02 \\
$uvw1$ &  1.48 & 1.51 \\
$uvm2$ &  1.71 & 1.69 \\
$uvw2$ &  1.72 & 1.73 \\
\hline
\enddata
\end{deluxetable}
\end{center}

\subsection{Photometric Stability}
\label{ss:variability}

The photometric uncertainty of any particular standard star's photometric measures is the combination
of the Poisson noise of the observation\footnote{With a small
correction in the UVOT arising from the finite number of CCD frames in an
observation (Kuin \& Rosen 2008), which can be
neglected for our sources.}, the uncertainty in the photometric zero
points and, in the case of aperture photometry, any variation in the
PSF.  As a photon-counting instrument, UVOT's read noise is irrelevant to the error budget.
In the case of UVOT, the first two sources
of uncertainty are quantified by our 
photometry pipeline and P08, respectively,
and included in Table \ref{t:photspace}.  The third -- variation of the PSF -- has been quantified by B10, along with other
small instrumental effects.  However, it can be independently checked from our standard star data.

The observations of our standard stars consist of 1-3 epochs of UVOT data.  Each of these epochs is, in turn, comprised
of many (mean of 30) independent UVOT exposures taken over multiple orbits of the {\it Swift} satellite.  The photometric
measures in Table
\ref{t:photspace} are taken from deep images produced by combining all the extant data.  However, the 10-45 independent UVOT
exposures that comprise each stacked
image allow us to check for any photometric zero point residuals in the data.
More importantly, the independent images allow a check on the photometric stability of the standard stars themselves.

We photometered the individual UVOT exposures using
the APPHOT package in IRAF, with apertures set to the 5\farcs0 optimal apertures specified in P08.
The raw photometry was corrected for coincidence loss using the formulation of P08 and calibrated using the transformations
of P08 and iterative matrix inversion techniques described in Siegel et al. (2002).  The iteration process
derives and corrects for exposure-to-exposure zero point residuals within each photometric passband,
with residuals measured from the mean zero point.

\begin{center}
\begin{figure}[ht!]
\includegraphics[scale=0.75]{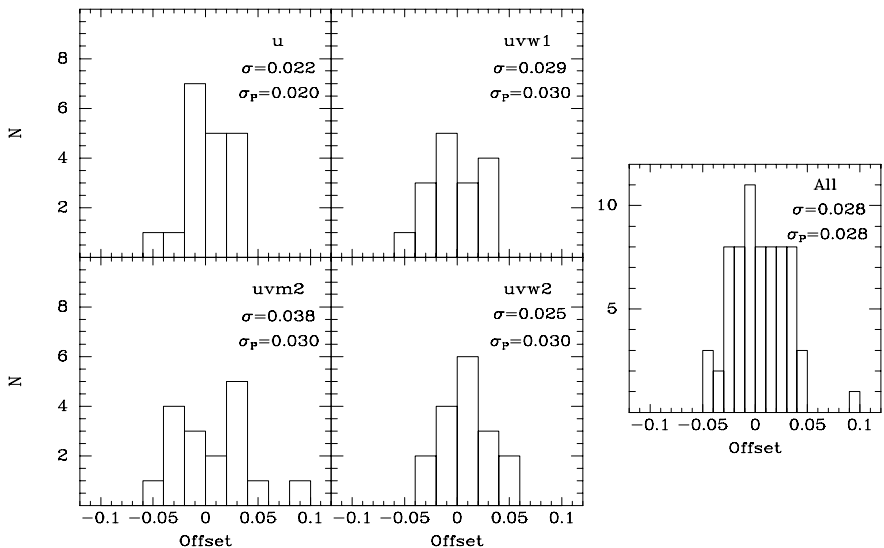}
\figcaption[f2.eps]{The distribution of zero point offsets from the mean frame for each individual UVOT 
image of SDSS J002806.49+010112.2.  The data are broken down by filter and then combined in the right-hand panel.
The distribution of the offsets is roughly Gaussian with typical $\sigma$ measures
of 0.02-0.04 magnitudes, similar in scale to the calibration uncertainty given in P08 (listed as $\sigma_P$).\label{f:oneoff}}
\end{figure}
\end{center}

Figure \ref{f:oneoff} shows a typical result of our investigation -- the distribution of exposure-to-exposure photometric
zero point residuals for SDSS J002806.49+010112.2.  The distribution is roughly Gaussian with a typical zero
point dispersion of 0.02-0.04 magnitudes.  This dispersion is comparable
to the scale of the instrumental effects quantified in B10.

\begin{center}
\begin{figure}[ht!]
\includegraphics[scale=0.75]{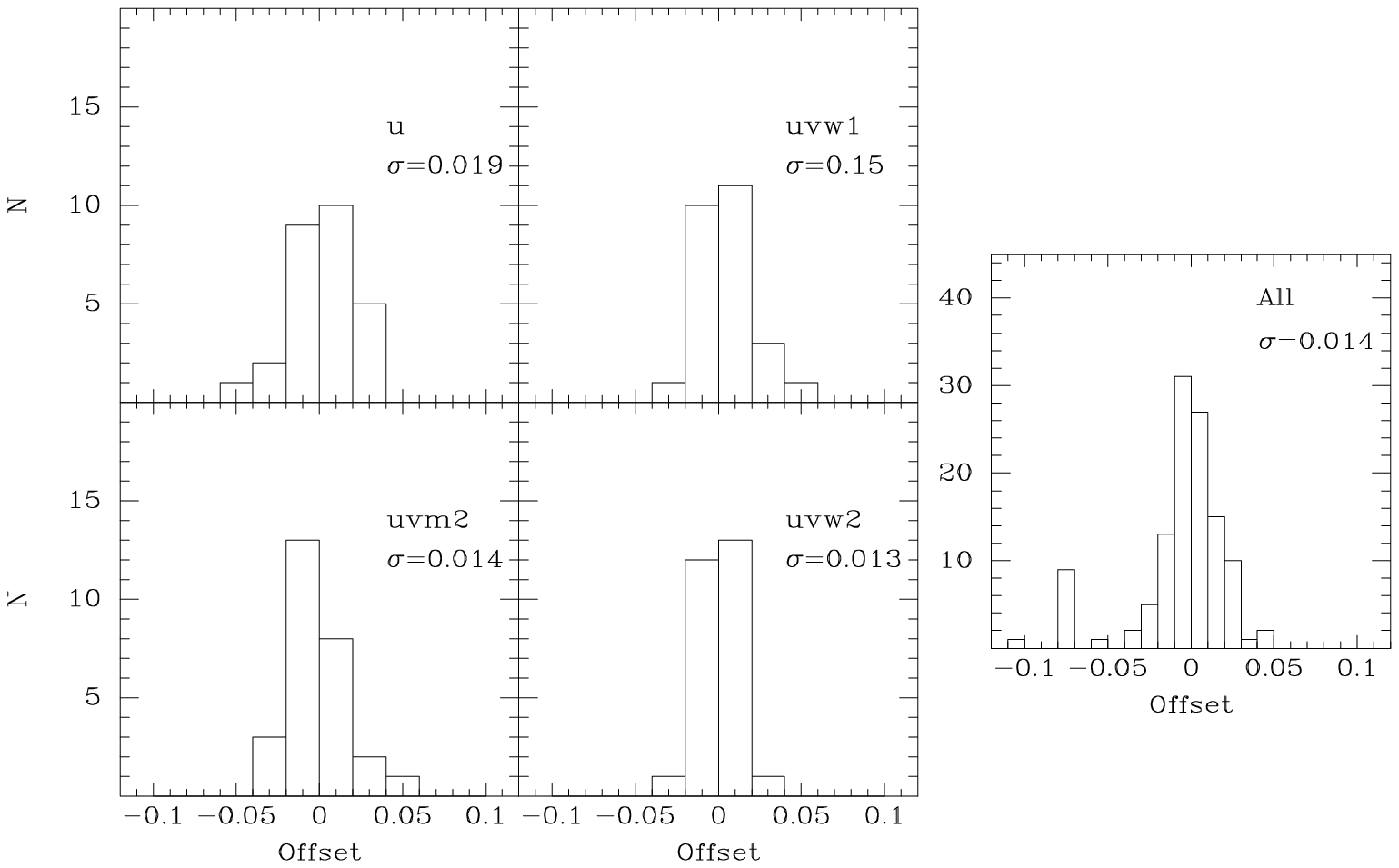}
\figcaption[f3.eps]{The distribution of zero point offsets of the combined UVOT epochs
from the mean for all eleven of our new standards.  The data are broken down by filter and
then combined in the right-hand panel.
The distribution of the offsets is roughly Gaussian with typical $\sigma$ measures
of 0.01-0.02 magnitudes\label{f:allframes}}
\end{figure}
\end{center}

The individual UVOT exposures, however, are shallow and have few stars with which to make comparisons
(2-60, with a median of 7).  With such small numbers of comparison stars, a single bad measure could dramatically
alter the measured residuals.  To improve the statistics, we 
combined the UVOT images from each epoch separately and photometered them using the techniques described above.
In this case, exposure time was no longer constant across the stacked image owing to the roll and pointing
uncertainty of the spacecraft, 
but was easily corrected from the exposure maps produced
by the UVOT reduction pipeline.  The dispersion of these epoch-to-epoch photometric residuals was calculated from a much deeper sample of
10-163 (median 30) common
stars and shows a much clearer Gaussian pattern with a dispersion of .01-.02 magnitudes (Figure \ref{f:allframes}).  The reduction
in the residual dispersion is consistent with having averaged out by subsampling some of the PSF variability in the combination.  We expect
that further improvement of the UVOT pipeline or attention to the systematics quantified by B10 will further reduce or eliminate these residual
effects.

\begin{center}
\begin{figure}[ht!]
\includegraphics[scale=0.75]{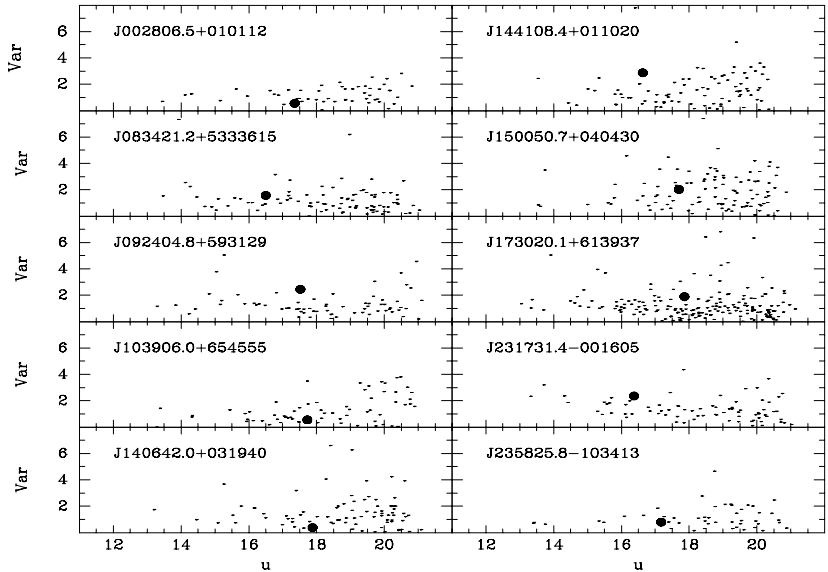}
\figcaption[f4.eps]{The ratio of observed scatter to measurement uncertainty of stars in the UV standard fields.  Large points are our new UV
standards.  The measures clump close to 1.0 with some notable
variable stars at high ratios.  None of our white dwarfs 
show significant long-term variability over the months-long timescale of the UVOT observations.  SDSS J134430.11+032423.2 is not plotted since it
had a single epoch of observation.\label{f:variability}}
\end{figure}
\end{center}

Removing these small zero point residuals using the aforementioned iterative calibration allows a more precise check on the 
variability of our new UV standards.
Figure \ref{f:variability} shows the ratio of measured photometric scatter to measurement uncertainty as a function of magnitude, after the correction
for the zero point residuals.  This measure is, essentially, the $\chi^2$ of a constant magnitude fit to the data.
While a few stars have high variability measures, the majority are clumped at low values, with a mean variability
index of 1.41 and a 90\% interval between 0.4 and 3.1.  This would be consistent with some residual zero point systematic error in the photometric
measures inflating the ratio over its expected value of 1.0.  The white dwarf stars have a mean variability index of 1.55 with a maximum of 2.87, well within
the bulk of field star distribution.  The variability indices are listed in the final column of Table \ref{t:photspace}.

Within the limits
measured by our UVOT program, our standard stars appear to be photometrically stable.  Further monitoring, to
measure any potential  variation over year-long timescales, is recommended.

As a further check on the photometric stability of the stars, we have examined the SDSS photometry of two stars which 
fall within the SDSS Southern Stripe (Stripe 82), a section of the DR4 which has been repeatedly observed,
yielding data of 1\% precision, half the more typical 2\% precision
of SDSS data (Ivezi{\'c} et al. 2007).  We confined our
analysis to data taken
before MJD 53400 (12:00 pm January 29, 2005 UT).  Beyond that date, the photometric measures are more
dense but include many data taken under non-photometric conditions.

\begin{center}
\begin{figure}[ht!]
\includegraphics[scale=0.75]{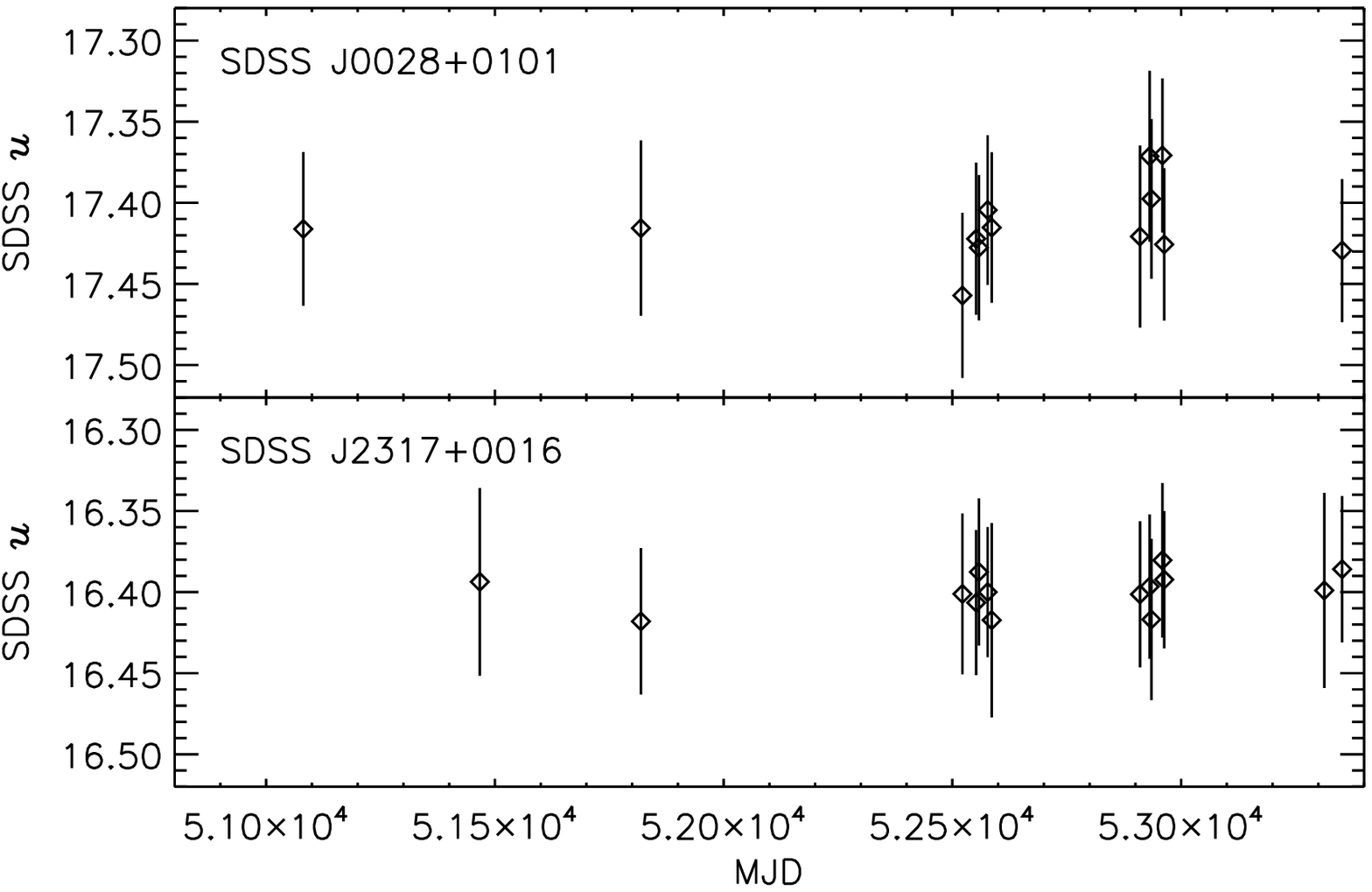}
\figcaption[f5.eps]{Photometric measures of two standard stars in the SDSS Southern Stripe over 5-6 years.\label{f:stripe82}}
\end{figure}
\end{center}

Figure \ref{f:stripe82} shows the measured photometry and no variability is seen.  The variability indices
are all significantly less than 1.0, indicating excellent photometric stability over the 5-6 year time scale of
the observational set.  We note that SDSS J231731.36-001604.9, which has a variability index in the Swift data of 2.35, shows miniscule
variability in the more extensive Sloan data.

As a final check, we took advantage of the simultaneous X-ray observations of our targets produced by the XRT.  While our white dwarf stars
are too faint and cold to produce noticeable X-ray flux, an X-ray signal could be produced in the (unlikely) event that one were
an accreting X-ray binary.  After running the automated analysis of Evans et al. (2009), however, we find no X-ray source
at the position of any of our standard stars.

\section{Comparison to Spectral Models}
\label{s:models}

With the suitability and photometric stability of our standards assured, we can
now constrain the underlying white dwarf spectral model and test the ability
of the models to reproduce the observed photometry.  A reliable model spectrum
could be used for calibrating other UV telescopes
regardless of whether their filter response
curves resemble those of UVOT or GALEX.  While Eisenstein et al. and AP09 have fit
SDSS white dwarf parameters from optical and near-infrared spectra,
our UV photometry provides additional leverage on the effective
temperatures of the white dwarfs because the UV samples a much more sensitive portion of the white dwarf
spectrum.

The complexity of modeling the photometry is greatly reduced by our limiting
of the candidate list to DA stars
without evidence of magnetic fields, metal lines or companions.  The
key parameters of the models are $T_{eff}$, $\log g$ and $A_V$, the $V$-band foreground extinction.

\begin{center}
\begin{figure}[ht!]
\includegraphics[scale=0.6,angle=90]{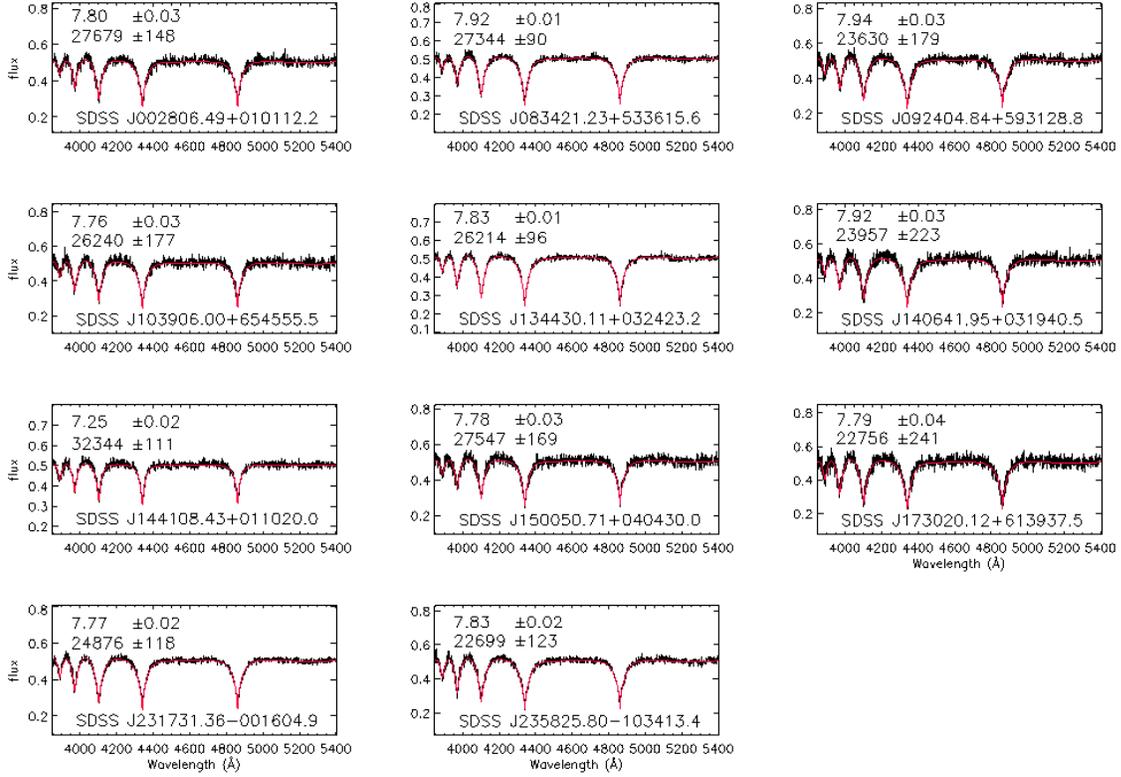}
\figcaption[f6.eps]{Model fits to the continuums-subtracted SDSS spectra, following the prescription of AP09.  As in AP09,
the parameters of the fits given in Table \ref{t:specfit} differ slightly from those shown in the figure.\label{f:specfit}}
\end{figure}
\end{center}

As a preliminary step, we refit the continuum-corrected SDSS spectra using the methods outlined in AP09.  This was done primarily to
better constrain $\log g$, which our photometry proved to be relatively insensitive to.  The parameters of the
purely spectral fits are given in Table \ref{t:specfit} and the continuum-corrected fits are
shown in Figure \ref{f:specfit}.  They are similar to those of Eisenstein et al.
with the notable exception of SDSS J092404.84+593128.8, for which we find a lower gravity.  In all cases, the $\chi^2$ of the fit is less than 1.0.

\begin{center}
\begin{deluxetable}{lccc}
\tablewidth{0 pt}
\tablecaption{Spectral Fits to Continuum-Corrected SDSS Spectra\label{t:specfit}}
\tablehead{
\colhead{Name} &
\colhead{$T_{eff}$} &
\colhead{$\log g$} &
\colhead{$\chi^2$}}
\startdata
\hline
SDSS J002806.49+010112.2 & 27679 (148) & 7.804 (0.025) & 0.63 \\
SDSS J083421.23+533615.6 & 27344  (90) & 7.916 (0.015) & 0.73 \\
SDSS J092404.84+593128.8 & 23630 (179) & 7.943 (0.025) & 0.62 \\
SDSS J103906.00+654555.5 & 26240 (177) & 7.762 (0.026) & 0.68 \\
SDSS J134430.11+032423.2 & 26214  (96) & 7.828 (0.014) & 0.51 \\
SDSS J140641.95+031940.5 & 23957 (223) & 7.924 (0.030) & 0.72 \\
SDSS J144108.43+011020.0 & 32344 (111) & 7.254 (0.024) & 0.56 \\
SDSS J150050.71+040430.0 & 27547 (169) & 7.777 (0.028) & 0.65 \\
SDSS J173020.12+613937.5 & 22756 (241) & 7.791 (0.035) & 0.68 \\
SDSS J231731.36-001604.9 & 24876 (118) & 7.768 (0.016) & 0.58 \\
SDSS J235825.80-103413.4 & 22699 (123) & 7.833 (0.018) & 0.67 \\
\hline
\enddata
\end{deluxetable}
\end{center}

Before fitting the photometry,
we corrected the models for extinction by simply taking the $V$-band extinction values ($A_V$) from the reddening maps of \cite{SFD} 
listed in Table 1 and applying the extinction curve give in \citet{Pei92} to the model spectra.  This technique is essentially an inversion of the
"extinction without standards" method outlined in Fitzpatrick \& Massa (2005), which combines photometry with spectral models to derive
a UV extinction curve.  In this case, we used photometry and an assumed extinction curve to constrain the spectral models.

Attempts were made to fit $A_V$ directly from the photometric measures by
producing multiple families of models with different foreground
extinction values and UV extinction laws.  However, we found the fitted $A_V$ values to be both imprecise (typical
fitting uncertainty was 0.1 magnitudes -- an uncertainty larger than the total \cite{SFD} measures) and co-variant
with the fitted $T_{eff}$.

It is possible that using the full \cite{SFD} values over-corrects the models for extinction.  The \citet{SFD}
values are for the full dust column and have been shown to be possibly over-estimated (see, e.g., Cambresy et al. 2005).
However, our white dwarfs are typically at distances of 150-250 pc.  Given an exponential
dust scale height of 134 pc (Marshall et al. 2006), it is likely that we are viewing the white dwarfs through 70-85\% of
the total dust along the line of sight.  The difference between this and the full \citet{SFD} values is less than 0.01 magnitudes.

The significance of a 0.01 magnitude uncertainty in the foreground extinction is small.
Analysis of our models indicates that changing the assumed reddening by 0.02 magnitude would
alter the derived temperatures by, on average, 100 degrees, with over-estimates of reddening producing over-estimates of
temperature and under-estimates being similarly covariant.  Thus, the effect of reddening
uncertainty on our model fitting
is expected to be less than that of the photometric uncertainties.

With revised $\log g$ values and $A_V$ values, we turned to constraining the white dwarf temperatures based
purely on the photometry.  We calculated a grid of pure hydrogen non-LTE models using Version 204 of Tlusty
(Lanz \& Hubeny, 1995), including the quasi-molecular satellites of Lyman $\alpha$ and Lyman $\beta$.  These 
models do not include the new Stark broadening calculations of Tremblay \&
Bergeron (2009), which could shift the derived $T_{eff}$ by as much as +1500 K.

A grid of model spectra was calculated for $T_{eff} = 20000$, 22500,
25000, 27500, 30000 K  and 35000 K and $\log g= 7.0$, 7.5, 8.0, 8.5 and 9.0.  We linearly interpolated
these spectra to a resolution of 100 K and 0.05 in $\log g$.  We then reddened the models based on the foreground
reddening and Pei dust curve.  At each temperature and gravity combination, synthetic magnitudes were then calculated by convolving
the synthetic spectrum with the most recent filter response curves for UVOT, GALEX and SDSS using the formulation:

\begin{equation}
m = -2.5 log \left( \frac{\int d(log \lambda) \lambda^2 f_{\lambda} S_{\lambda}}{\int d(log \lambda) S_{\lambda}}
\right) + C
\end{equation}

where $f_{\lambda}$ is the model spectrum flux and $S_{\lambda}$ is the system throughput per unit wavelength.  $C$ is the zero
point magnitude.

Observed magnitudes were compared to model magnitudes using
\begin{equation}
\chi^2 = \displaystyle\sum_i \frac{(m_i - m_{model,i} + c)^2}{\sigma_i^2}
\label{eq:chisq}
\end{equation}
\noindent where $m_i$ is the observed magnitude,
$m_{model,i}$ is the model magnitude, and $\sigma_i$ is the photometric error for
in the $i$th filter.  The summation is over all 11 filters shown in Tables \ref{t:photsdss} and \ref{t:photspace}
and described in \S\ref{ss:photometry}.

The value $c$ is a
normalization constant which matches the observed and model spectra to
the same level.  The optimal value of $c$ can be analytically
determined by setting $\partial \chi^2/\partial c = 0$.  Solving for
$c$ yields
\begin{equation}
c = \frac{\left ( \displaystyle\sum_i \frac{m_{model,i}}{\sigma_i^2}-
 \displaystyle\sum_i \frac{m_i}{\sigma_i^2}\right )}
 {\displaystyle\sum_i 1/\sigma_i^2}
\end{equation}
The value of $c$ was calculated independently for each combination of observed and model magnitudes and use to calculate
and minimize $\chi^2$.

To quantify the non-linear effect of photometric uncertainty upon our model fits, we performed
a Monte Carlo simulation on the data.  The photometric measures were perturbed by the photometric
errors and $T_{eff}$ was refit
using Equation \ref{eq:chisq}.  A general description
of the Monte Carlo technique can be found in \citet{Press92}.
Given the relative brightness of our candidate stars,
the photometric errors were dominated by systematic zero point errors and
not the random Poisson errors for all stars in all bands.  Zero point
errors were modeled as uniformly distributed with end points set the photometric
uncertainties.  For each star, this process was
repeated for 10000 Monte Carlo realizations.  The resulting
distributions of $T_{eff}$ was used to determine the best
values from the median and 90\% error bars.

\begin{center}
\begin{figure}[ht!]
\includegraphics[scale=0.9]{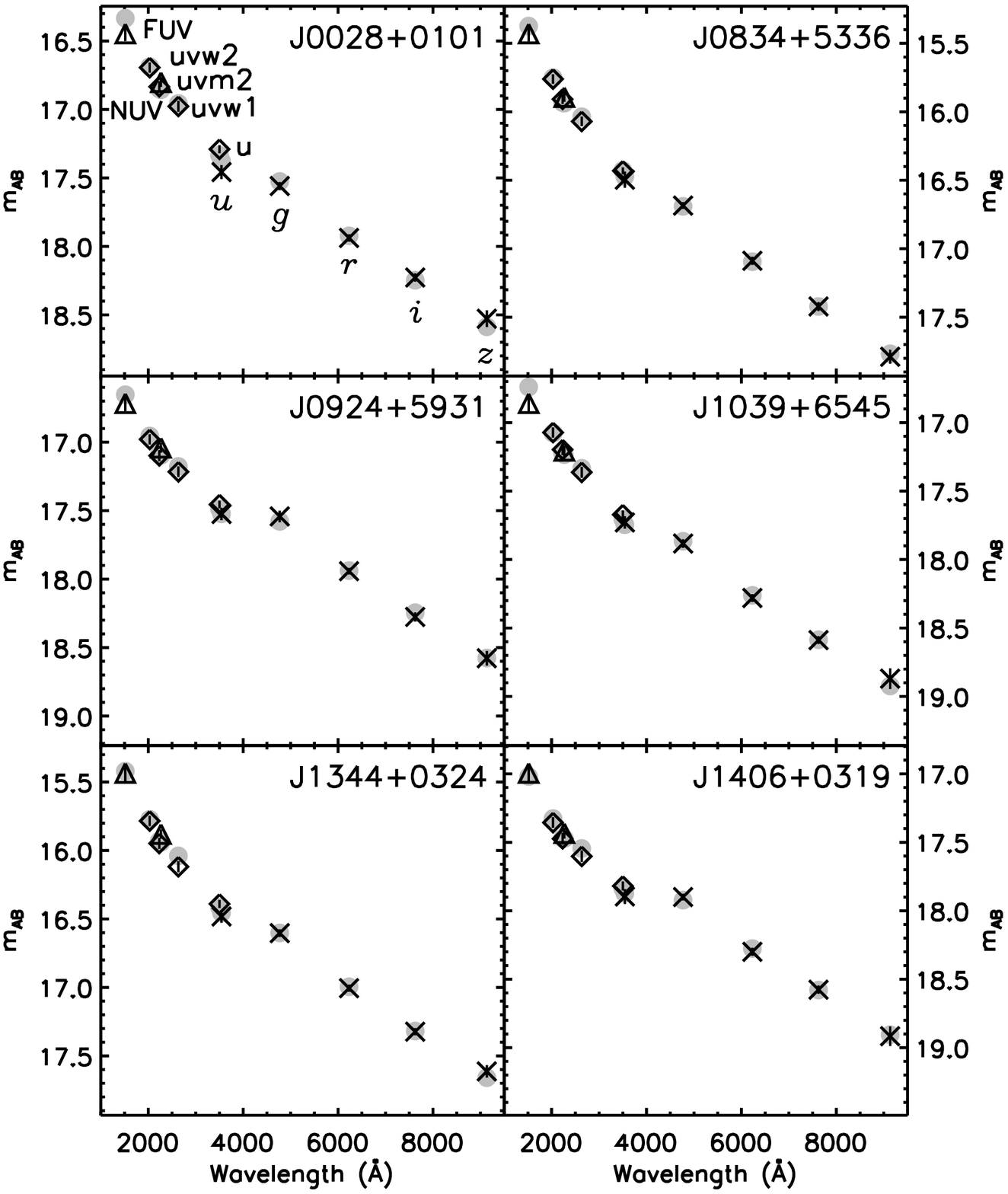}
\figcaption[f7.eps]{Comparison of the white dwarf photometry to the predicted magnitudes of the model atmospheres.
The grey dots represent the model; X's, diamonds and triangles represent SDSS, UVOT and GALEX photometry,
respectively. Individual filters are labeled in the first panel. \label{f:phot_plot}}
\end{figure}
\end{center}

\begin{center}
\begin{figure}[ht!]
\includegraphics[scale=0.9]{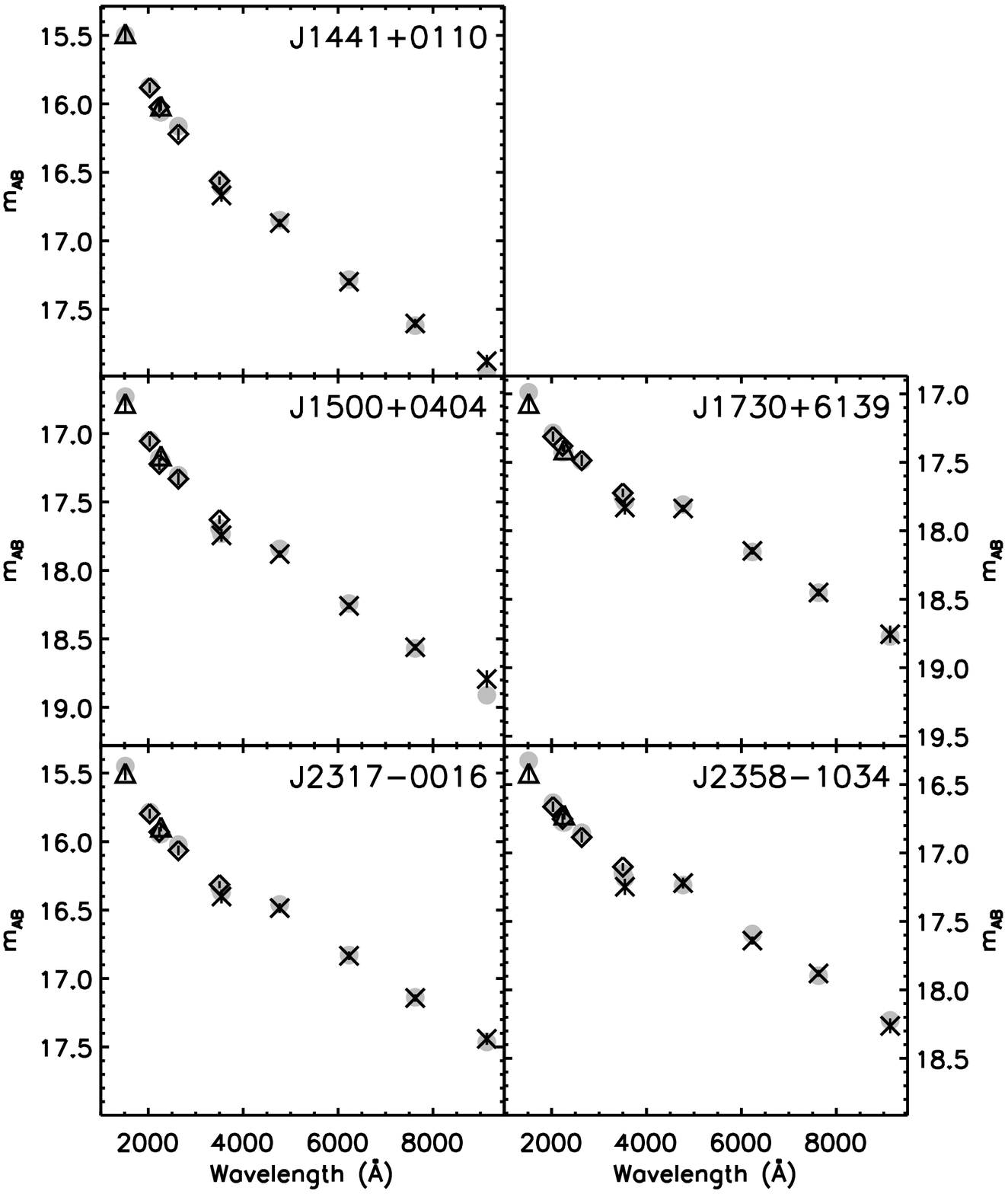}
\figcaption[f8.eps]{Comparison of the white dwarf photometry to the predicted magnitudes of the model atmospheres.
The grey dots represent the model; X's, diamonds and triangles represent SDSS, UVOT and GALEX photometry,
respectively. \label{f:phot_plotB}}
\end{figure}
\end{center}

\begin{center}
\begin{figure}[ht!]
\includegraphics[scale=0.7]{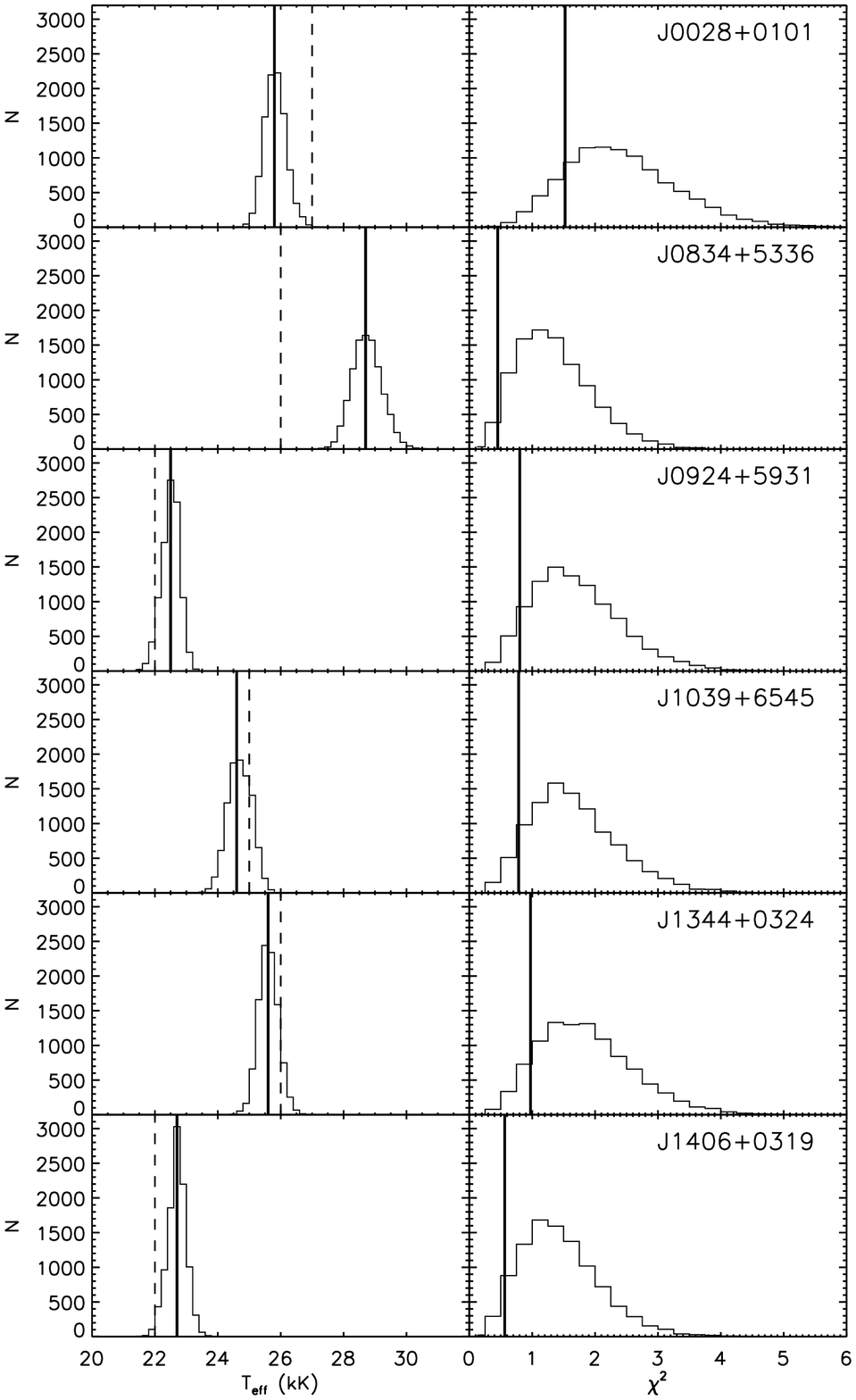}
\figcaption[f9.eps]{Monte Carlo simulation of white dwarf parameters for our UV standards.
The histogram indicates the distribution of model fits after perturbing the photometry with zero point offsets
drawn from a uniform distribution set to the stated random and systematic
uncertainties in the photometric measures.
The dashed line represents the $T_{eff}$ reported in Eisenstein et al.; the solid lines
represent our best fit from the photometry.\label{f:mc_A}}
\end{figure}
\end{center}

\begin{center}
\begin{figure}[ht!]
\includegraphics[scale=0.7]{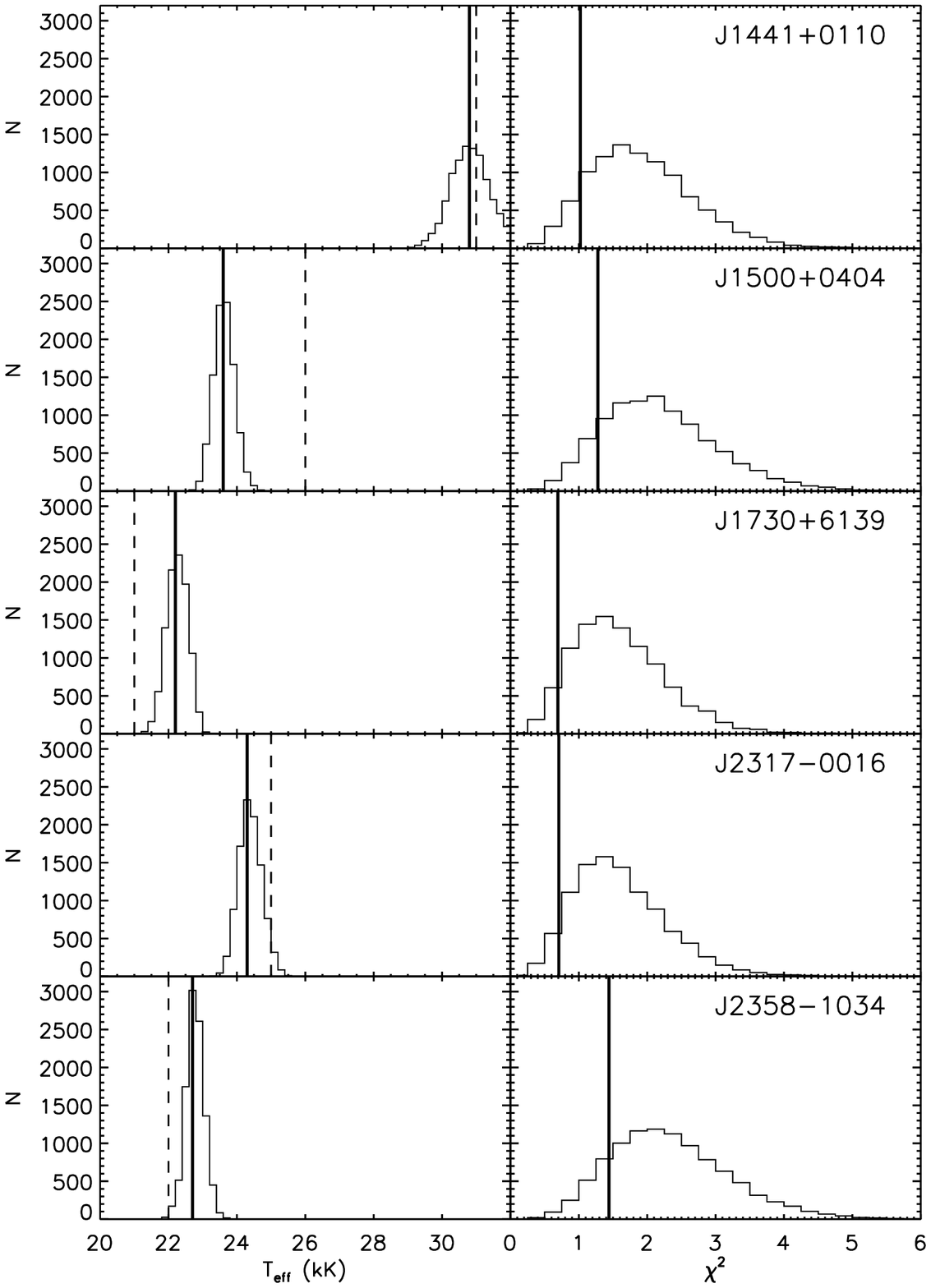}
\figcaption[f10.eps]{Monte Carlo simulation of white dwarf parameters for our UV standards.
The histogram indicates the distribution of model fits after perturbing the photometry with zero point offsets
drawn from a uniform distribution set to the stated random and systematic
uncertainties in the photometric measures.
The dashed line represents the $T_{eff}$ reported in Eisenstein et al.; the solid lines
represent our best fit from the photometry.\label{f:mc_B}}
\end{figure}
\end{center}

\begin{center}
\begin{figure}[ht!]
\includegraphics[scale=0.7]{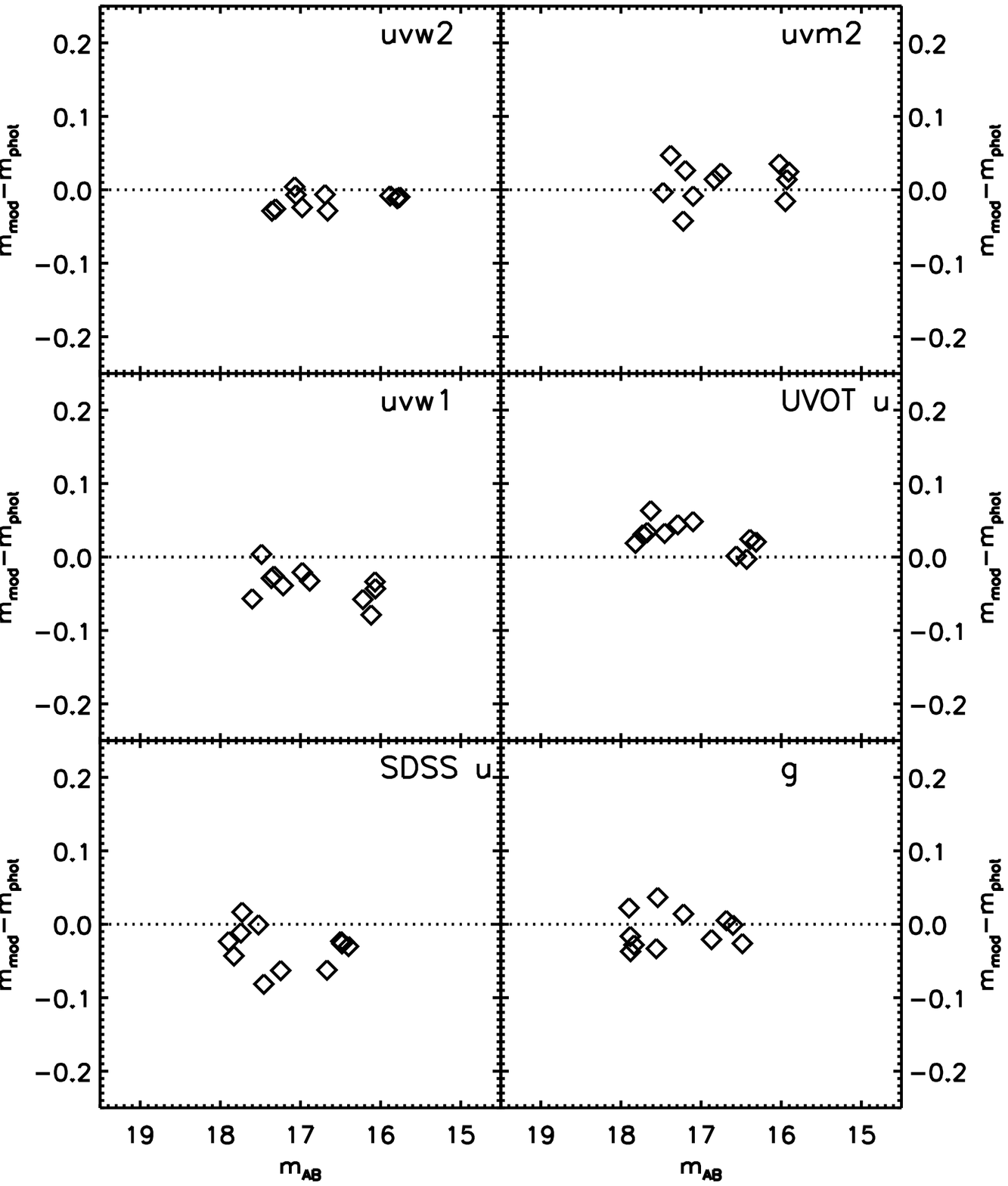}
\figcaption[f11.eps]{The photometric residuals of fitted white dwarf models to measured photometry broken down by filter.\label{f:resid_A}}
\end{figure}
\end{center}

\begin{center}
\begin{figure}[ht!]
\includegraphics[scale=0.7]{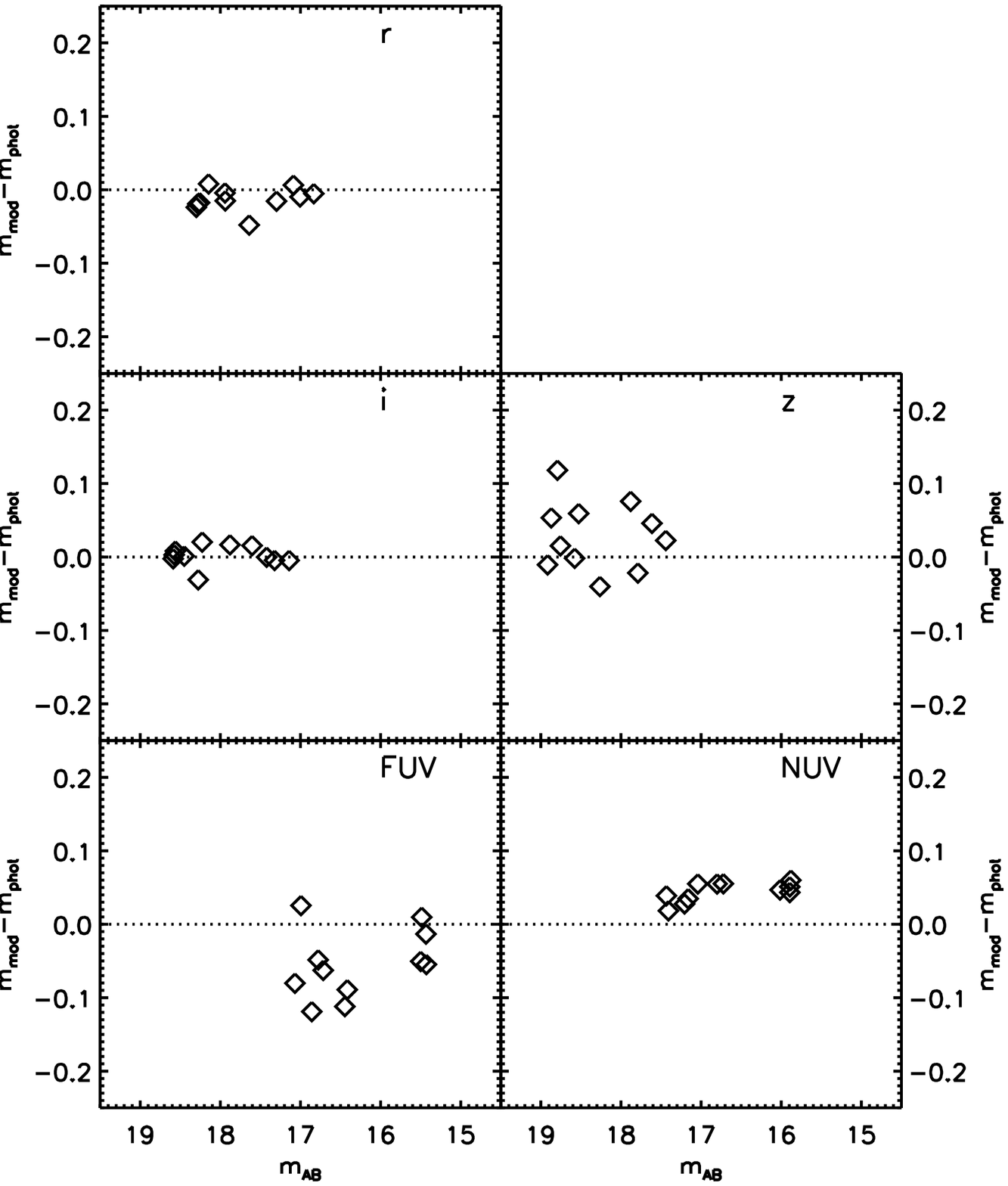}
\figcaption[f12.eps]{The photometric residuals of fitted white dwarf models to measured photometry broken down by filter.\label{f:resid_B}}
\end{figure}
\end{center}

The best-fit model parameters are given in Table \ref{t:modteff}. The model
fits and Monte Carlo error simulations are shown in figures
\ref{f:phot_plot}-\ref{f:mc_B}.  For all of our dwarf stars, we find a 
favored $T_{eff}$ that accurately reproduces the observed photometric
measures across all eleven passbands.  No stars has a reduced $\chi^2$ greater than 1.51 and the mean $\chi^2$ is 0.92, indicating
that the residuals are within the photometric uncertainties.

\begin{center}
\begin{figure}[ht!]
\includegraphics[scale=0.7]{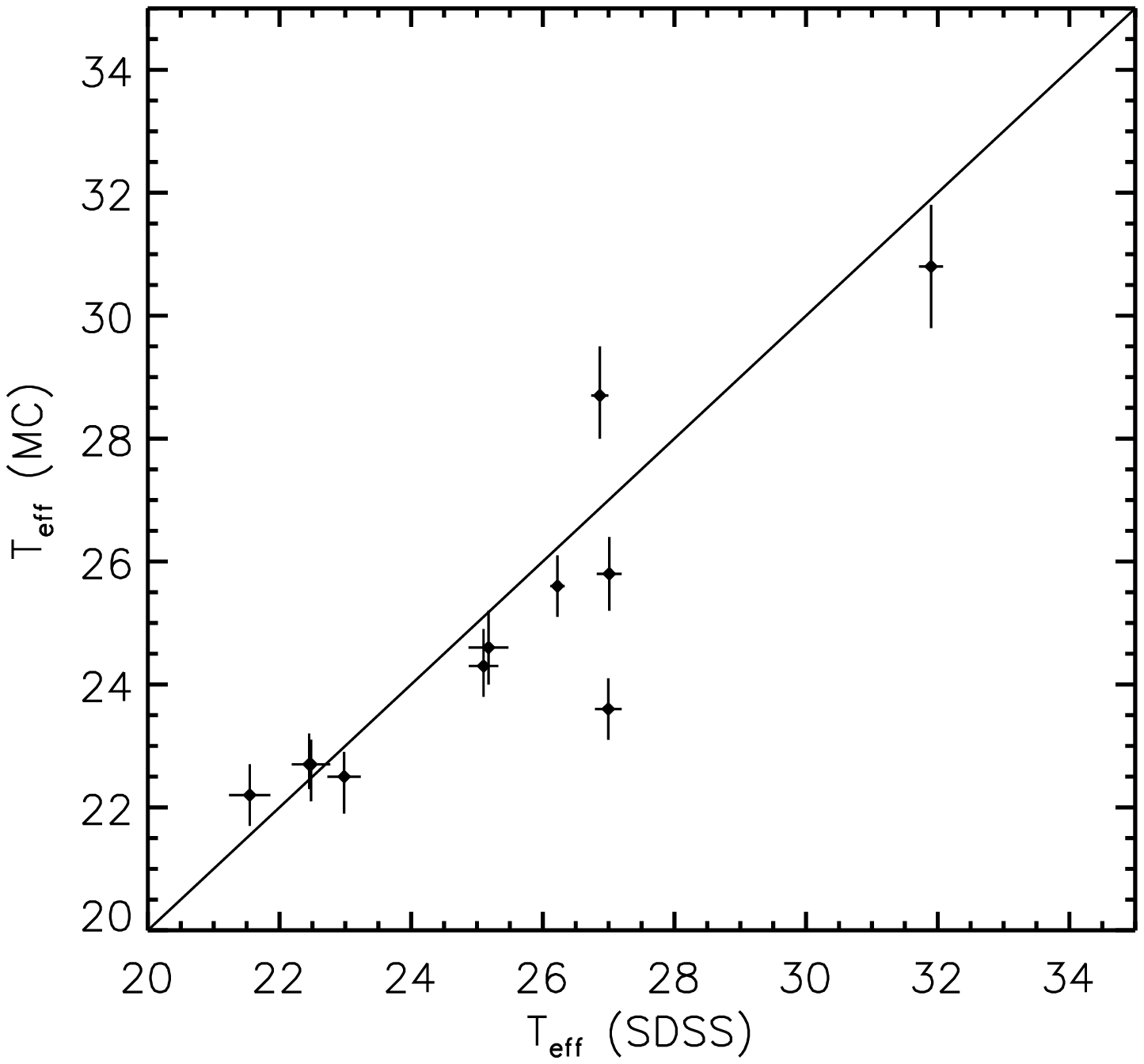}
\figcaption[f13.eps]{A comparison of the SDSS effective temperatures from Eisenstein et al. against those
derived from our spectral fitting.  The solid line indicates unity.\label{f:sdssmctemp}}
\end{figure}
\end{center}

Figures \ref{f:resid_A} and \ref{f:resid_B} show the photometric residuals ($\Delta_m=m_{model}-m_{data}$) for all
of our stars in each filter while Table \ref{t:photresid} list the mean, median and dispersion of the residuals in comparison to the systematic
uncertainties in the photometric zero points ($\sigma_{sys}$) for each filter.
In almost all cases, the mean and median $\Delta_m$ are within one or two $\sigma_{sys}$ of zero, indicating that our models
agree with the measured photometry within the 
zero point uncertainties compiled in Tables \ref{t:photsdss} and \ref{t:photspace}.
The mean $\frac{|<\Delta_m>|}{\sigma_{sys}}$ is 0.99, indicating excellent agreement
between the models and data.  This confirms that the spectra
can be used by future investigations to extrapolate predicted magnitudes in other passbands.  It also
confirms the quality and consistency of the most recent UVOT calibration.

\begin{center}
\begin{deluxetable}{lcccc}
\tablewidth{0 pt}
\tabletypesize{\small}
\tablecaption{Best Fit Effective Temperatures of UV Standards\label{t:modteff}}
\tablehead{
\colhead{Name} &
\colhead{$T_{eff}$ 90\% Lower Bound} &
\colhead{$T_{eff}$ Best Fit} &
\colhead{$T_{eff}$ 90\% Upper Bound} &
\colhead{$\chi^2_{red}$}}
\startdata
\hline
SDSS J002806.49+010112.2 & 25200 & 26100 & 26400 &  1.51\\
SDSS J083421.23+533615.6 & 28000 & 28500 & 29500 &  0.45\\
SDSS J092404.84+593128.8 & 22500 & 22500 & 22900 &  0.72\\
SDSS J103906.00+654555.5 & 24800 & 24800 & 25200 &  0.78\\
SDSS J134430.11+032423.2 & 25600 & 25600 & 26100 &  0.97\\
SDSS J140641.95+031940.5 & 22700 & 22700 & 23100 &  0.58\\
SDSS J144108.43+011020.0 & 30800 & 30800 & 31800 &  1.01\\
SDSS J150050.71+040430.0 & 23300 & 23300 & 24100 &  1.28\\
SDSS J173020.12+613937.5 & 22300 & 22300 & 22700 &  0.68\\
SDSS J231731.36-001604.9 & 24300 & 24300 & 24900 &  0.71\\
SDSS J235825.80-103413.4 & 23000 & 23000 & 23200 &  1.44\\

\hline
\enddata
\end{deluxetable}
\end{center}

\begin{center}
\begin{deluxetable}{ccccc}
\tablewidth{0 pt}
\tablecaption{Residuals of Fits to Photometry\label{t:photresid}}
\tablehead{
\colhead{Filter} &
\colhead{$<\Delta_m>$} &
\colhead{median ($\Delta_m$)} &
\colhead{$\sigma_{\Delta_m}$} &
\colhead{$\sigma_{sys}$}}
\startdata
\hline
  FUV  & -0.054 & -0.055 & 0.047 & 0.05\\
  NUV  &  0.044 &  0.047 & 0.013 & 0.03\\
 $uvw2$  & -0.014 & -0.011 & 0.011 & 0.03\\
 $uvm2$  &  0.010 &  0.014 & 0.026 & 0.03\\
 $uvw1$  & -0.038 & -0.034 & 0.022 & 0.03\\
 $u$ (UVOT) &  0.029 &  0.031 & 0.019 & 0.02\\
 $u$ (SDSS) & -0.032 & -0.026 & 0.029 & 0.03\\
    $g$  & -0.008 & -0.016 & 0.025 & 0.01\\
    $r$  & -0.013 & -0.015 & 0.015 & 0.01\\
    $i$  &  0.002 & -0.001 & 0.014 & 0.01\\
    $z$  &  0.029 &  0.023 & 0.047 & 0.02\\
\hline
\enddata
\end{deluxetable}
\end{center}

Figure \ref{f:sdssmctemp} compares the Eisenstein et al. (2006) 
effective temperatures against those derived in our model fitting.  While the values roughly track each
other, we find our temperatures average slightly lower than those given in Eisenstein et al.
A similar result was found in AP09, who found the Eisenstein et al. temperatures to be a few percent
lower than theirs up to 30,000 degrees and slightly warmer at higher temperatures.

\begin{center}
\begin{deluxetable}{cccccccccccc}
\tabletypesize{\tiny}
\tablewidth{0 pt}
\tablecaption{Model Flux\tablenotemark{a} for WD Standards\label{t:electromod}}
\tablehead{
\colhead{Wavelength (\AA)} &
\colhead{J002806.49+010112.2} &
\colhead{J083421.23+533615.6} &
\colhead{J092404.84+593128.8} &
\colhead{J103906.00+654555.5} &
\colhead{J134430.11+032423.2} &
\colhead{J140641.95+031940.5} &
\colhead{J144108.43+011020.0} &
\colhead{J150050.71+040430.0} &
\colhead{J173020.12+613937.5} &
\colhead{J231731.36-001604.9} &
\colhead{J235825.80-103413.4}}
\startdata
\hline
 1300.0 & 2.0081E-13 & 4.7217E-13 & 1.4265E-13 & 1.3793E-13 & 4.6467E-13 & 1.0135E-13 & 4.3014E-13 & 1.3937E-13 & 1.0286E-13 & 4.4074E-13 & 1.9253E-13\\
 1300.5 & 2.0057E-13 & 4.7166E-13 & 1.4253E-13 & 1.3776E-13 & 4.6414E-13 & 1.0126E-13 & 4.2971E-13 & 1.3920E-13 & 1.0276E-13 & 4.4024E-13 & 1.9235E-13\\
 1301.0 & 2.0037E-13 & 4.7121E-13 & 1.4243E-13 & 1.3763E-13 & 4.6369E-13 & 1.0119E-13 & 4.2929E-13 & 1.3907E-13 & 1.0270E-13 & 4.3986E-13 & 1.9221E-13\\
 1301.5 & 2.0017E-13 & 4.7067E-13 & 1.4233E-13 & 1.3750E-13 & 4.6323E-13 & 1.0112E-13 & 4.2880E-13 & 1.3894E-13 & 1.0263E-13 & 4.3949E-13 & 1.9208E-13\\
 1302.0 & 1.9997E-13 & 4.7021E-13 & 1.4223E-13 & 1.3737E-13 & 4.6278E-13 & 1.0105E-13 & 4.2835E-13 & 1.3881E-13 & 1.0257E-13 & 4.3912E-13 & 1.9194E-13\\
 1302.5 & 1.9980E-13 & 4.6976E-13 & 1.4213E-13 & 1.3727E-13 & 4.6239E-13 & 1.0098E-13 & 4.2794E-13 & 1.3870E-13 & 1.0251E-13 & 4.3884E-13 & 1.9181E-13\\
 1303.0 & 1.9968E-13 & 4.6939E-13 & 1.4214E-13 & 1.3717E-13 & 4.6211E-13 & 1.0097E-13 & 4.2753E-13 & 1.3862E-13 & 1.0250E-13 & 4.3860E-13 & 1.9178E-13\\
 1303.5 & 1.9948E-13 & 4.6894E-13 & 1.4204E-13 & 1.3704E-13 & 4.6166E-13 & 1.0090E-13 & 4.2708E-13 & 1.3849E-13 & 1.0244E-13 & 4.3822E-13 & 1.9164E-13\\
 1304.0 & 1.9933E-13 & 4.6848E-13 & 1.4204E-13 & 1.3695E-13 & 4.6134E-13 & 1.0090E-13 & 4.2667E-13 & 1.3842E-13 & 1.0242E-13 & 4.3798E-13 & 1.9161E-13\\
 1304.5 & 1.9913E-13 & 4.6803E-13 & 1.4194E-13 & 1.3681E-13 & 4.6089E-13 & 1.0083E-13 & 4.2626E-13 & 1.3829E-13 & 1.0236E-13 & 4.3761E-13 & 1.9147E-13\\
\hline
\enddata
\tablenotetext{a}{Flux in $erg cm^{-2} s^{-1} A^{-1}$}
\end{deluxetable}
\end{center}

The fitted spectral models are included in the electronic version of the {\it Astrophysical Journal} (Table \ref{t:electromod}).  These can be compared to
measured spectra or convolved with filter functions using the procedure outlined above to predict standard stellar magnitudes in non-UVOT
passbands.  However, given uncertainties
in any model fitting and the likelihood of future updates to the TLUSTY code, it may be advisable for
future calibrations to use our published photometry and fit parameters as a starting point for a refined exploration of these new standard
stars.

\section{Conclusions}
\label{s:conc}

We have created a catalog of 11 faint DA white dwarf ultraviolet standards for use with space-based UV detectors.  Our stars
have been carefully chosen for simple modeling, low extinction, a lack of nearby stellar companions, even distribution across
the sky and faintness that will avert problems of saturation or coincidence loss.  In combination with SDSS and GALEX, we 
provide precise photometry for these stars from the NIR to the FUV.
Checks from both UVOT and SDSS data show that our white dwarf stars are photometrically stable.
When combined with recent sample of spectrophotometric
standards recommended by AP09, up to 20 new UV standard stars are now potentially available to the community, all of which have published
SDSS photometry and eleven of which now have published NUV and FUV photometry.

We have fit both the SDSS spectra and measured photometry of our standard stars with 
relatively simple white dwarf stellar atmospheric models.  We find that these
models provide strong and consistent constraints on the properties of the white dwarf stars, reproducing the photometric measures to within the 1-5\%
uncertainty of their photometric zero points.  This indicates outstanding suitably of our standard stars for
calibration of future missions that may not share the filter set of UVOT and GALEX through the
use of simple white dwarf atmospheric models.

Of our eleven stars, we do not find any that are of poor or limited quality.  The mean reduced $\chi^2$ of the model fits to spectra and 
photometry are 0.64 and 0.92, respectively, indicating excellent agreement between models and data.  Measurement of stellar variability
shows some stars to have moderately elevated
photometric scatter.  However, this scatter is within the distribution of stable field stars and does not appear in the more extensive SDSS Stripe 82 data.
We do, however, recommend further monitoring to ensure that the stars are photometrically stable.

The two most significant uncertainties in our standard stars are (1) the systematic uncertainty in the photometric zero points of GALEX
and UVOT, which limit the models to reproducing the photometry within the 1-5\% uncertainty of the zero points; (2) the 
previously known uncertainty in the UV extinction law.
These systematic uncertainties dominate over our random errors.  We are engaged in effort to better understand the UV extinction
curve in Galactic dust.  We also recommend further investigation to improve the calibration of both GALEX and UVOT.
In combination, these two endeavors -- dust properties and calibration -- would 
enhance the precision of our faint UV standard to better than 1\%, which would both improve the capabilities of UVOT and
help other missions to explore the ultraviolet range of the spectrum.

The good agreement between model, spectra and photometry is indicative of the outstanding suitability of DA White Dwarfs as
UV standard stars.  As the models and especially our understanding of the UV properties of the dust improve, our ability to characterize the
white dwarf properties will also improve.  This will allow further investigation into more distant and/or more reddened white dwarfs
to improve our understanding of these stellar relics.

\acknowledgements

The authors acknowledge support in the form of GALEX grant NNX08AK62G and sponsorship at PSU by
NASA contract NAS5-00136. We thank I. Hubeny for assistance with the TLUSTY software and the anonymous
referee for useful comments.

\bibliographystyle{apj}

\end{document}